\documentclass[12pt]{article}
\usepackage{amssymb}

\usepackage{graphicx}
\usepackage{amsmath}

\input{tcilatex}

\begin{document}

\title{THE MEANING OF DIMENSIONS}
\date{}
\author{\textbf{Paul S. Wesson}$^{1,2}$}
\maketitle

\begin{enumerate}
\item \bigskip Department of Physics, University of Waterloo, \newline
Waterloo, Ontario N2L 3G1, Canada

\item Gravity-Probe B, Hansen Physics Labs., Stanford University, \newline
Stanford, California 94305, U.S.A.
\end{enumerate}

\begin{center}
\bigskip
\end{center}

A paper for the book edited by Professor V. Petkov, Dept. of Physics,
Concordia University, Montreal, Quebec H3G 1M8, Canada.\newline
[vpetkov@alcor.concordia.ca]

\ \ \ \ \newline

Correspondence: Mail to address (1) above. Phone = (519) 885-1211 x2215,
email (not for publication) = pswastro@hotmail.com\newpage

\begin{center}
{\Large THE MEANING OF DIMENSIONS}
\end{center}

{\large Abstract}

We review the current status of dimensions, as the result of a long and
controversial history that includes input from philosophy and physics. Our
conclusion is that they are subjective but essential concepts which provide
a kind of book-keeping device, their number increasing as required by
advances in physics. The world almost certainly has more than the four
dimensions of space and time, but the introduction of the fifth and higher
dimensions requires a careful approach wherein known results are embedded
and new ones are couched in the most productive manner.

\section{\protect\underline{Introduction}}

\ \ \ \ \ Dimensions are both primitive concepts that provide a framework
for mechanics and sophisticated devices that can be used to construct
unified field theories. Thus the ordinary space of our perceptions $\left(
xyz\right) $ and the subjective notion of time $\left( t\right) $ provide
the labels with which to describe Newtonian mechanics, or with the
introduction of the speed of light to form an extra coordinate $\left(
ct\right) $ the mechanics of 4D Einstein relativity. But used in the
abstract, they also provide a means of extending general relativity in
accordance with certain physical principles, as in 10D supersymmetry. As
part of the endeavour to unify gravity with the interactions of particle
physics, there has recently been an explosion of interest in manifolds with
higher dimensions. Much of this work is algebraic in nature, and has been
reviewed elsewhere (see below). Therefore, to provide some balance and
direction, we will concentrate here on fundamentals and attempt to come to
an understanding of the meaning of dimensions.

Our main conclusion, based on 35 years of consideration, will be that
dimensions are basically inventions, which have to be chosen with skill if
they are to be profitable in application to physics

This view may seem strange to some workers, but is not new. It is implicit
in the extensive writings on philosophy and physics by the great astronomer
Eddington, and has been made explicit by his followers, who include the
writer. This view is conformable, it should be noted, with algebraic proofs
and other mathematical results on many-dimensional manifolds, such as those
of the classical geometer Campbell, whose embedding theorem has been
recently rediscovered and applied by several workers to modern unified-field
theory. Indeed, a proper understanding of the meaning of dimensions involves
both history and modern physics.

There is a large literature on dimensions; but it would be inappropriate to
go into details here, and we instead list some key works. The main
philosophical/physical ones are those by Barrow (1981), Barrow and Tipler
(1986), Eddington (1935, 1939), Halpern (2004), Kilmister (1994), McCrea and
Rees (1983), Petley (1985), Price and French (2004) and Wesson (1978, 1992).
The main algebraic/mathematical works are those by Campbell (1926), Green et
al. (1987), Gubser and Lykken (2004), Seahra and Wesson (2003), Szabo
(2004), Wesson (1999, 2006) and West (1986). These contain extensive
bibliographies, and we will quote freely from them in what follows.

The plan of this paper is as follows. Section 2 outlines the view of our
group, that dimensions are inventions whose geometrical usefulness for
physics involves a well-judged use of the fundamental constants. This rests
on work by Eddington, Campbell and others; so in Sections 3 and 4 we give
accounts of the main philosophical and algebraic results (respectively) due
to these men, in a modern context. Section 5 is a summary, where we restate
our view that the utility of dimensions in physics owes at least as much to
skill as to symbolism. We aim to be pedagogical rather than pedantic, and
hope that the reader will take our comments in the spirit of learning rather
than lecture.

\section{\protect\underline{Dimensions and Fundamental Constants}}

\ \ \ \ \ Minkowski made a penetrating contribution to special relativity
and our view of mechanics when by the simple identification of $x^{4}\equiv
ct$ he put time on the same footing as the coordinates $\left(
x^{123}=xyz\right) $ of the ordinary space of our perceptions. \ Einstein
took an even more important step when he made the Principle of Covariance
one of the pillars of general relativity, showing that the 4 coordinates
traditionally used in mechanics can be altered and even mixed, producing an
account of physical phenomena which is independent of the labels by which we
choose to describe them. \ These issues are nowadays taken for granted; but
a little reflection shows that insofar as the coordinates are the labels of
the dimensions, the latter are themselves flexible in nature.

Einstein was in his later years also preoccupied with the manner in which we
describe matter. \ His original formulation of general relativity involved a
match between a purely geometrical object we now call the Einstein tensor $%
\left( G_{\alpha \beta },\,\alpha \text{\ and }\beta =0,123\text{\ for }%
t,xyz\right) $, and an object which depends on the properties of matter
which is known as the energy-momentum (or stress-energy) tensor ($T_{\alpha
\beta },$ which\ contains quantities like the ordinary density $\rho $\ and
pressure $p$ of matter). \ The coefficient necessary to turn this
correspondence into an equation is (in suitable units) $8\pi G\diagup c^{4}$%
, where $G$ is the gravitational constant. \ Hence, Einstein's field
equations, $G_{\alpha \beta }=\left( 8G\diagup c^{4}\right) T_{\alpha \beta
} $, which are an excellent description of gravitating matter. \ In writing
these equations, it is common to read them from left to right, so that the
geometry of 4D spacetime is governed by the matter it contains. \ However,
this split is artificial. \ Einstein himself realized this, and sought
(unsuccessfully) for some way to turn the ``base wood'' of $T_{\alpha \beta
} $ into the ``marble'' of $G_{\alpha \beta }$. \ His aim, simply put, was
to geometrize all of mechanical physics - the matter as well as the fields.

A potential way to geometrize the physics of gravity and electromagnetism
was suggested in 1920 by Kaluza, who added a fifth dimension to Einstein's
general relativity. \ Kaluza showed in essence that the apparently empty 5D
field equations $R_{AB}=0\left( A,B=0,123,4\right) $ in terms of the Ricci
tensor, contain Einstein's equations for gravity and Maxwell's equations for
electromagnetism. \ Einstein, after some thought, endorsed this step. \
However, in the 1920s quantum mechanics was gaining a foothold in
theoretical physics, and in the 1930s there was a vast expansion of interest
in this area, at the expense of general relativity. \ This explains why
there was such a high degree of attention to the proposal of Klein, who in
1926 suggested that the fifth dimension of Kaluza ought to have a closed
topology (i.e., a circle), in order to explain the fundamental quantum of
electric charge $\left( l\right) $. \ Klein's argument actually related this
gravity to the momentum in the extra dimension, but in so doing introduced
the fundamental unit of action $\left( h\right) $which is now known as
Planck's constant. \ However, despite the appeal of Klein's idea, it was
destined for failure. \ There are several technical reasons for this, but it
is sufficient to note here that the crude 5D gravity/quantum theory of
Kaluza/Klein implied a basic role for the mass quantum $\left( h\,c\diagup
G\right) ^{1/2}$. \ This is of order 10$^{-5}$ g, and does \underline{not}
play a dominant role in the spectrum of masses observed in the real
universe. \ (In more modern terms, the so-called hierarchy problem is
centred on the fact that observed particle masses are far less than the
Planck mass, or any other mass derivable from a tower of states where this
is a basic unit.) \ Thus, we see in retrospect that the Klein modification
of the Kaluza scheme was a dead-end. \ This does not, though, imply that
there is anything wrong with the basic proposition, which follows from the
work of Einstein and Kaluza, that matter can be geometrized with the aid of
the fundamental constants. \ As a simple example, an astrophysicist
presented with a problem involving a gravitationally-dominated cloud of
density $\rho $ will automatically note that the free-fall or dynamical
timescale is the inverse square root of $G\rho $. \ This tells him
immediately about the expected evolution of the cloud. \ Alternatively,
instead of taking the density as the relevant physical quantity, we can form
the length $\left( c^{2}\diagup G\rho \right) ^{1/2}$ and obtain an
equivalent description of the physics in terms of a geometrical quantity.

The above simple outline, of how physical quantities can be combined with
the fundamental constants to form geometrical quantities such as lengths,
can be much developed and put on a systematic basis (Wesson 1999). \ The
result is induced-matter theory, or as some workers prefer to call it,
space-time-matter theory. \ The philosophical basis of the theory is to
realize Einstein's dream of unifying geometry and matter (see above). \ The
mathematical basis of it is Campbell's theorem, which ensures an embedding
of 4D general relativity with sources in a 5D theory whose field equations
are apparently empty (see below). \ That is, the Einstein equations $%
G_{\alpha \beta }=\left( 8\pi G\diagup c^{4}\right) T_{\alpha \beta }\left(
\alpha ,\beta =0,123\right) $ are embedded perfectly in the Ricci-flat
equations $R_{AB}=0\left( A,B=0,123,4\right) $. \ The point is, in simple
terms, that we use the fifth dimension to model matter.

An alternative version of 5D gravity, which is mathematically similar, is
membrane theory. \ In this, gravity propagates freely in 5D, into the
``bulk''; but the interactions of particles are confined to a hypersurface
or the ``brane''. \ It has been shown by Ponce de Leon and others that both
the field equations and the dynamical equations are effectively the same in
both theories. \ The only difference is that whereas induced-matter theory
treats all 5 dimensions as equivalent, membrane theory makes spacetime a
special (singular) hypersurface. \ For induced-matter theory, particles can
wander away from the hypersurface at a slow rate governed by the
cosmological constant; whereas for membrane theory, particles are confined
to the hypersurface by an exponential force governed by the cosmological
constant. \ Both versions of 5D general relativity are in agreement with
observations. \ The choice between them is largely philosophical: Are we
living in a universe where the fifth dimension is ``open'', or are we living
an existence where we are ``stuck'' to a particular slice of 5D manifold?

Certainly, the fundamental constants available to us at the present stage in
the development of physics allow us to geometrize matter in terms of 
\underline{one} extra dimension. \ Insofar as mechanics involves the basic
physical quantities of mass, length and time, it is apparent that any code
for the geometrization of mass will serve the purpose of extending 4D
spacetime to a 5D space-time-mass manifold (the theory is covariant). \
However, not all parametizations are equally convenient, in regard to
returning known 4D physics from a 5D definition of ``distances'' or metric.
\ Thus, the ``canonical'' metric has attracted much attention. \ In it, the
line element is augmented by a flat extra dimension, while its 4D part is
multiplied by a quadratic factor (the corresponding metric is membrane
theory involves an exponential factor, as noted above). \ The physics flows
from this factor, which is $\left( l\diagup L\right) ^{2}$where $x^{4}=l$
and $L$ is a constant which by comparison with the 4D Einstein metric means $%
L=\left( 3\diagup \Lambda \right) ^{1/2}$ where $\Lambda $\ is the
cosmological constant. \ In this way, we weld ordinary mechanics to
cosmology, with the identification $x^{4}=l=Gm\diagup c^{2}$ where $m$ is
the rest mass of a macroscopic object. \ If, on the other hand, we wish to
study microscopic phenomena, the simple coordinate transformation $%
l\rightarrow L^{2}\diagup l$ gives us a quantum (as opposed to classical)
description of rest mass via $x^{4}=h\diagup mc$. \ In other words, the
large and small scales are accommodated by choices of coordinates which
utilize the available fundamental constants, labelling the mass either by
Schwarzchild radius or by the Compton wavelength.

It is not difficult to see how to extend the above approach to higher
dimensions. \ However, skill is needed here. \ For example, electric charge
can either be incorporated into 5D (along the lines\ originally proposed by
Kaluza and Klein), or treated as a sixth dimension (with coordinate $%
x_{q}\equiv \left( G\diagup c^{4}\right) ^{1/2}q$ where $q$ is the charge,
as studied by Fukui and others). \ A possible resolution of technical
problems like this is to ``fill up'' the parameter space of the
lowest-dimensional realistic model (in this case 5D), before moving to a
higher dimension. \ As regards other kinds of ``charges'' associated with
particle physics, they should be geometrized and then treated as coordinates
in the matching $N$-dimensional manifold. \ In this regard, as we have
emphasized, there are choices to be made about how best to put the physics
into correspondence with the algebra. \ For example, in supersymmetry, every
integral-spin boson is matched with a half-integral-spin fermion, in order
to cancel off the enormous vacuum or zero-point fields which would otherwise
occur. \ Now, it is a theorem that any curved energy-full \ solution of the
4D Einstein field equations can be embedded on a flat and energy-free 10D
manifold. \ (This is basically a result of counting the degrees of freedom
in the relevant sets of equations: see Section 4 below). \ This is the
simplest motivation known to the writer for supersymmetry. \ However, it is
possible in certain cases that the condition of zero energy can be
accomplished in a space of less than 10 dimensions, given a skillful choice
of parameters.

We as physicists have chosen geometry as the currently best way to deal with
macroscopic and microscopic mechanics; and while there are theorems which
deal with the question of how to embed the 4D world of our senses in
higher-dimensional manifolds, the choice of the latter requires intuition
and skill.

\section{\protect\underline{Eddington and His Legacy}}

\ \ \ \ \ In studying dimensions and fundamental constants over several
decades, the writer has come to realize that much modern work on these
topics has its roots in the views of Arthur Stanley Eddington (1882-1944;
for a recent interdisciplinary review of his contribution to physics and
philosophy, see the conference notes edited by Price and French, 2004). \ He
was primarily an astronomer, but with a gift for the pithy quote. \ For
example: ``We are bits of stellar matter that got cold by accident, bits of
a star gone wrong''. \ However, Eddington also thought deeply about more
basic subjects, particularly the way in which science is done, and was of
the opinion that much of physics is subjective, insofar as we necessarily
filter data about the external world through our human-based senses. \ Hence
the oft-repeated quote: ``To put the conclusion crudely-the stuff of the
world is mind-stuff''. \ The purpose of the present section is to give a
short and informal account of Eddington's views, and thereby alert workers
in fundamental physics to his influence.

This was primarily through a series of non-technical books and his personal
contacts with a series of great scientists who followed his lead. \ These
include Dirac, Hoyle and McCrea. \ In the preceding section, we noted that
while it is possible to add an arbitrary number of extra dimensions to
relativity as an exercise in mathematics, we need to use the fundamental
constants to identify their relevance to physics. \ (We are here talking
primarily about the speed of light $c$, the gravitational constant $G$ and
Planck's constant of action $h$, which on division by 2$\pi $ also provides
the quantum of spin angular momentum.) \ To appreciate Eddington's legacy,
we note that his writings contain the first logical account of the large
dimensionless numbers which occur in cosmology, thereby presaging what Dirac
would later formalize as the Large Numbers Hypothesis. \ This consists
basically in the assertion that large numbers of order 10$^{40}$ are in fact
equal, which leads among other consequences to the expectation that $G$ is
variable over the age of the universe (see Wesson 1978). \ This possibility
is now discussed in the context of field theory in $N>4$ dimensions, where
the dynamics of the higher-dimensional manifold implies that the coupling
constants (like $G$) in 4D are changing functions of the spacetime
coordinates (Wesson 1999). \ One also finds in Eddington's works some very
insightful, if controversial, comments about the so-called fundamental
constants. \ These appear to have influenced Hoyle, who argued that the $%
c^{2}$ in the common relativistic expression $\left(
c^{2}t^{2}-x^{2}-y^{2}-z^{2}\right) $ should not be there, because ``there
is no more logical reason for using a different time unit than there would
be for measuring $x,\,y,\,z$ \ in different units''. \ The same influence
seems to have acted on McCrea, who regarded $c$, $G$ and $h$ as ``conversion
constants and nothing more''. \ These comments are in agreement with the
view advanced in Section 2, namely that the fundamental constants are
parameters which can be used to change the physical units of material
quantities to lengths, enabling them to be given a geometrical description.
\ There is a corollary of this view which is pertinent to several modern
versions of higher-dimensional physics. \ Whatever the size of the manifold,
the equations of the related physics are homogeneous in their physical units 
$\left( M,\,L,\,T\right) $ so they can always be regarded as equalities
involving \underline{dimensionless} parameters. \ It makes sense to consider
the possible (say) time variation of such parameters; but it makes no sense
to argue that the component \underline{dimensionalful} quantities are
variable. \ To paraphrase Dicke: Physics basically consists of the
comparison of dimensionless parameters at different points in the manifold.

Views like this \underline{still} raise the hackles of certain physicists
who have not analysed the problem at a deep level. \ Eddington, in
particular, was severely criticized by both physicists and philosophers when
he presented his opinions in the 1930s. \ Fortunately, many workers - as a
result of their studies of unified field theory - came to a sympathetic
understanding of Eddington's opinions in the 1990s. \ However, there is an
interesting question of psychology involved here.

Plato tells us of an artisan whose products are the result of experience and
skill and meet with the praise of his public for many years. \ However, in
later times he suddenly produces a work which is stridently opposed to
tradition and incurs widespread criticism. \ Has the artisan suffered some
delusion, or has he broken through to an art form so novel that his
pedestrian-minded customers cannot appreciate or understand it?

Eddington spent the first part of his academic career doing well-regarded
research on stars and other aspects of conventional astronomy. \ He then
showed great insight and mathematical ability in his study of the then-new
subject of general relativity. \ In his later years, however, he delved into
the arcane topic of the dimensionless numbers of physics, attempting to
derive them from an approach which combined elements of pure reason and
mathematics. \ This approach figures significantly in his book \underline{%
Relativity Theory of Protons and Electrons}, and in the much-studied
posthumous volume \underline{Fundamental Theory}. \ The approach fits
naturally into his philosophy of science, which latter argued that many
results in physics are the result of \underline{how} we do science, rather
than direct discoveries about the external world (which, however, he
admitted). \ Jeffreys succeeded Eddington to the Plumian Chair at Cambridge,
but was a modest man more interested in geophysics and the formation of the
solar system than the speculative subject of cosmology. \ Nevertheless, he
developed what at the time was a fundamental approach to the theory of
probability, and applied his skills to a statistical analysis of Eddington's
results. \ The conclusion was surprising: according to Jeffreys' analysis of
the uncertainties in the underlying data which Eddington had used to
construct his account of the basic physical parameters, the results agreed
with the data \underline{better than they ought to have done}. \ This raised
the suspicion that Eddington had ``cooked'' the results. \ This author spent
the summer of 1970 in Cambridge, having written (during the preceding summer
break from undergraduate studies at the University of London) a paper on
geophysics which appealed to Jeffreys. \ We discussed, among other things,
the status of Eddington's results. \ Jeffreys had great respect for
Eddington's abilities, but was of the opinion that his predecessor had 
\underline{unwittingly} put subjective elements into his approach which
accounted for their unreasonable degree of perfection. \ The writer pointed
out that there was another possible explanation: that Eddington was in fact
right in his belief that the results of physics were derivable from first
principles, and that his approach was compatible with a more profound theory
which yet awaits discovery.

\section{\protect\underline{Campbell and His Theorem}}

\ \ \ \ \ Whatever the form of a new theory which unifies gravity with the
forces of particle physics, there is a consensus that it will involve extra
dimensions. \ In Section 2, we considered mainly the 5D approach, which by
the modern names of induced-matter and membrane theory is essentially old
Kaluza-Klein theory without the stifling condition of compactification. \
The latter, wherein the extra dimension is ``rolled up'' to a very small
size, answers the question of why we do not ``see'' the fifth dimension. \
However, an equally valid answer to this is that we are constrained to live
close to a hypersurface, like an observer who walks across the surface of
the Earth without being directly aware of what lies beneath his feet. \ In
this interpretation, 5D general relativity must be regarded as a kind of new
standard. \ It is the simplest extension of Einstein's theory, and is widely
viewed as the low-energy limit of more sophisticated theories which
accommodate the internal symmetry groups of particle physics, as in 10D
supersymmetry, 11D supergravity and 26D string theory. \ There is, though,
no sancrosanct value of the dimensionality $N$. \ It has to be chosen with a
view to what physics is to be explained. \ (In this regard, St. Kalitzin
many years ago considered $N\rightarrow \infty $.) \ All this understood,
however, there is a practical issue which needs to be addressed and is
common to all higher-$N$ theories: How do we embed a space of dimension $N$
in one of dimension $\left( N+1\right) $? \ This is of particular relevance
to the embedding of 4D Einstein theory in 5D Kaluza-Klein theory. \ We will
consider this issue in the present section, under the rubric of Campbell's
theorem. \ While it is central and apparently simple, it turns out to have a
rather long history with some novel implications.

John Edward Campbell was a professor of mathematics at Oxford whose book ``A
Course of Differential Geometry'' was published posthumously in 1926. \ The
book is basically a set of lecture notes on the algebraic properties of $N$D
Riemannian manifolds, and the question of embeddings is treated in the
latter part (notably chapters 12 and 14). \ However, what is nowadays called
Campbell's theorem is there only sketched. \ He had intended to add a
chapter dealing with the relation between abstract spaces and Einstein's
theory of general relativity (which was then a recent addition to physics),
but died before he could complete it. \ The book was compiled with the aid
of Campbell's colleague, E.B. Elliot, but while accurate is certainly
incomplete.

The problem of embedding an $N$D (pseudo-) Riemannian manifold in a
Ricci-flat space of one higher dimension was taken up again by Magaard. \ He
essentially proved the theorem in his Ph.D. thesis of 1963. \ This and
subsequent extensions of the theorem have been discussed by Seahra and
Wesson (2003), who start from the Gauss-Codazzi equations and consider an
alternative proof which can be applied to the induced-matter and membrane
theories mentioned above.

The rediscovery of Campbell's theorem by physicists can be attributed
largely to the work of Tavakol and coworkers. \ They wrote a series of
articles in mid-1990s which showed a connection between the CM theorem and a
large body of earlier results by Wesson and coworkers (Wesson 1999). \ The
latter group had been using 5D geometry as originally introduced by Kaluza
and Klein to give a firm basis to the aforementioned idea of Einstein, who
wished to transpose the ``base-wood'' of the right-hand side of his field
equations into the ``marble'' of the left-hand side. \ That an effective or
induced 4D energy-momentum tensor $T_{\alpha \beta }$ can be obtained from a
5D geometrical object such as the Ricci Tensor $R_{AB}$ is evident from a
consideration of the number of degrees of freedom involved in the problem
(see below). \ The only requirement is that the 5D metric tensor be left
general, and not be restricted by artificial constraints such as the
``cylinder'' condition imposed by Kaluza and Klein. \ Given a 5D line
element $dS^{2}=g_{AB}\left( x^{\gamma },l\right) dx^{A}dx^{B}\left(
A,B=t,xyz,l\right) $ it is then merely a question of algebra to show that
the equations $R_{AB}=0$ contain the ones $G_{\alpha \beta }=T_{\alpha \beta
}$ named after Einstein.  (In accordance with comments about the
non-fundamental nature of the constants, and common practice, we in this
section choose units which render $8\pi G\diagup c^{4}$ equal to unity.) \
Many exact solutions of $R_{AB}=0$ are now known (see Wesson 2006 for a
catalog). \ Of these, special mention should be made of the ``standard'' 5D
cosmological ones due to Ponce de Leon, and the 1-body and other solutions
in the ``canonical'' coordinates introduced by Mashhoon et al. \ It says
something about the divide between physics and mathematics, that the
connection between these solutions and the CM theorem was only made later,
by the aforementioned work of Tavakol et al. \ Incidentally, these workers
also pointed out the implications of the CM theorem for lower-dimensional $%
\left( N<4\right) $ gravity, which some researchers believe to be relevant
to the quantization of this force.

The CM theorem, which we will re-prove below, is a \underline{local}
embedding theorem. \ It cannot be pushed towards solving problems which are
the domain of (more difficult) \underline{global} embeddings. \ This implies
that the CM theorem should not be applied to initial-value problems or
situations involving singularities. \ It is a modest - but still very useful
- result, whose main implication is that we can gain a better understanding
of matter in 4D by looking at the field equations in 5D.

The CM theorem in succinct form says: Any analytic Riemannian space $%
V_{n}\left( s,t\right) $ can be locally embedded in a Ricci-flat Riemannian
space $V_{n+1}\left( s+1,t\right) $ or $V_{n+1}\left( s,t+1\right) $.

We are here using the convention that the ``small'' space has dimensionality 
$n$ with coordinates running 0 to $n-1$, while the ``large'' space has
dimensionality $n+1$ with coordinates running 0 to $n$. \ The total
dimensionality is $N=1+n$, and the main physical focus is on $N=5$.

The CM theorem provides a mathematical basis for the induced-matter theory,
wherein matter in 4D as described by Einstein's equations $G_{\alpha \beta
}=T_{\alpha \beta }$ is derived from apparent vacuum in 5D as described by
the Ricci-flat equations $R_{AB}=0$ (Wesson 1999, 2006). \ The main result
is that the latter set of relations satisfy the former set if 
\begin{eqnarray*}
T_{\alpha \beta } &=&\frac{\Phi _{,\alpha ;\beta }}{\Phi }-\frac{\varepsilon 
}{2\Phi ^{2}}\left\{ \frac{\Phi _{,4}g_{\alpha \beta ,4}}{\Phi }-g_{\alpha
\beta ,44}+g^{\lambda \mu }g_{\alpha \lambda ,4}g_{\beta \mu ,4}\right.  \\
&&\left. -\frac{g^{\mu \nu }g_{\mu \nu ,4}g_{\alpha \beta ,4}}{2}+\frac{%
g_{\alpha \beta }}{4}\left[ g_{,4}^{\mu \nu }g_{\mu \nu ,4}+\left( g^{\mu
\nu }g_{\mu \nu ,4}\right) ^{2}\right] \right\} \;\;\;\;\;.
\end{eqnarray*}%
Here the 5D line element is $dS^{2}=g_{\alpha \beta }\left( x^{\gamma
},l\right) dx^{\alpha }dx^{\beta }+\varepsilon \Phi ^{2}\left( x^{\gamma
},l\right) dl^{2}$, where $\varepsilon =\pm 1$, a comma denotes the ordinary
partial derivative and a semicolon denotes the ordinary 4D covariant
derivative. \ Nowadays, it is possible to prove Campbell's theorem using the
ADM formalism, whose lapse-and-shift technique has been applied extensively
to derive the energy of 5D solutions. \ It is also possible to elucidate the
connection between a smooth 5D manifold (as in induced-matter theory) and
one containing a singular surface (as in membrane theory). \ We now proceed
to give an ultra-brief account of this subject.

Consider an arbitrary manifold $\Sigma _{n}$ in a Ricci-flat space $V_{n+1}$%
. \ The embedding can be visualized by drawing a line to represent $\Sigma
_{n}$ in a surface, the normal vector $n^{A}$ to it satisfying $n\cdot
n\equiv n^{A}n_{A}=\varepsilon =\pm 1$. \ If $e_{\left( \alpha \right) }^{A}$
represents an appropriate basis and the extrinsic curvature of $\Sigma _{n}$
is $K_{\alpha \beta }$, the ADM constraints read 
\begin{eqnarray*}
G_{AB}n^{A}n^{B} &=&-\frac{1}{2}\left( \varepsilon R_{\alpha }^{\alpha
}+K_{\alpha \beta }K^{\alpha \beta }-K^{2}\right) =0 \\
G_{AB}e_{\left( \alpha \right) }^{A}n^{B} &=&K_{\alpha ;\beta }^{\beta
}-K_{,\alpha }=0\;\;\;\;\;.
\end{eqnarray*}%
These relations provide $1+n$ equations for the $2\times n\left( n+1\right)
\diagup 2$ quantities $g_{\alpha \beta }$, $K_{\alpha \beta }$. \ Given an
arbitrary geometry $g_{\alpha \beta }$ for $\Sigma _{n}$, the constraints
therefore form an under-determined system for $K_{\alpha \beta }$, so
infinitely many embeddings are possible.

This demonstration of Campbell's theorem can easily be extended to the case
where $V_{n+1}$ is a de Sitter space or anti-de Sitter space with an
explicit cosmological constant, as in some applications of brane theory. \
Depending on the application, the remaining $n\left( n+1\right) -\left(
n+1\right) =\left( n^{2}-1\right) $ degrees of freedom may be removed by
imposing initial conditions on the geometry, physical conditions on the
matter, or conditions on a boundary.

The last is relevant to brane theory with the $Z_{2}$ symmetry, where $%
dS^{2}=g_{\alpha \beta }\left( x^{\gamma },l\right) dx^{\alpha }dx^{\beta
}+\varepsilon dl^{2}$ with $g_{\alpha \beta }=g_{\alpha \beta }\left(
x^{\gamma },+l\right) $ for $l\geq 0$ and $g_{\alpha \beta }=g_{\alpha \beta
}\left( x^{\gamma },-l\right) $ for $l\leq 0$ in the bulk. \
Non-gravitational fields are confined to the brane at $l=0$, which is a
singular surface. \ Let the energy-momentum in the brane be represented by $%
\delta \left( l\right) S_{AB}$ (where $S_{AB}n^{A}=0$) and that in the bulk
by $T_{AB}$. \ Then the field equations read $G_{AB}=\kappa \left[ \delta
\left( l\right) S_{AB}+T_{AB}\right] $ where $\kappa $\ is a 5D coupling
constant. \ The extrinsic curvature discussed above changes across the brane
by an amount $\Delta _{\alpha \beta }\equiv K_{\alpha \beta }\left( \Sigma
_{l>0}\right) -K_{\alpha \beta }\left( \Sigma _{l<0}\right) $ which is given
by the Israel junction conditions. \ These imply 
\begin{equation*}
\Delta _{\alpha \beta }=-\kappa \left( S_{\alpha \beta }-\frac{1}{3}%
Sg_{\alpha \beta }\right) \;\;\;\;\;.
\end{equation*}%
But the $l=0$ plane is symmetric, so%
\begin{equation*}
K_{\alpha \beta }\left( \Sigma _{l>0}\right) =-K_{\alpha \beta }\left(
\Sigma _{l<0}\right) =-\frac{\kappa }{2}\left( S_{\alpha \beta }-\frac{1}{3}%
Sg_{\alpha \beta }\right) \;\;\;\;\;.
\end{equation*}%
This result can be used to evaluate the 4-tensor 
\begin{equation*}
P_{\alpha \beta }\equiv K_{\alpha \beta }-Kg_{\alpha \beta }=-\frac{\kappa }{%
2}S_{\alpha \beta }\;\;\;\;\;.
\end{equation*}%
However, $P_{\alpha \beta }$ is actually identical to the 4-tensor $\left(
g_{\alpha \beta ,4}-g_{\alpha \beta }g^{\mu \nu }g_{\mu \nu ,4}\right)
\diagup 2\Phi $ of induced-matter theory, where it figures in 4 of the 15
field equations $R_{AB}=0$ as $P_{\alpha ;\beta }^{\beta }=0$ (Wesson 1999).
\ That is, the conserved tensor $P_{\alpha \beta }$ of induced-matter theory
is essentially the same as the total energy-momentum tensor in $Z_{2}$%
-symmetric brane theory. \ Other correspondences can be established in a
similar fashion.

Thus while induced-matter theory and membrane theory are often presented as
alternatives, they are in fact the same thing, and from the viewpoint of
differential geometry both are rooted in the CM theorem. \ This theorem also
has the wider implication that, given the physics in a given manifold, we
can always derive the corresponding physics in a manifold of plus-or-minus
one dimension. \ In other words, Campbell's theorem provides a kind of
ladder which enables us to go up or down between manifolds of different
dimensionality.

\section{\protect\underline{Summary}}

Dimensions are a delightful subject with which to dally, but we should
remind ourselves that they need the cold scrutiny of common sense to be
useful. \ This means, among other things, that we should have a physical
identification of the extra coordinates, in order to understand the
implications of their associated dimensions. \ In 5D, we have seen that the
extra coordinate can profitably be related to rest mass, either as the
Schwarzschild radius or the Compton wavelength, in the classical and quantum
domains respectively. \ This implies that the fifth dimension is a scalar
field, which is presumably the classical analog of the Higgs field by which
particles acquire mass in quantum field theory. \ This interpretation
depends on a judicial use of the fundamental constants (Section 2). \ This
approach gives much to the work of Eddington, who delved deeply into the
meanings of the equations of physics (Section 3). \ Our usage of dimensions
also owes something to Campbell, whose theorem in its modern form shows how
to go between manifolds whose dimensionality differs by one (Section 3). \
Our conclusion is that while the use of dimensions may in some respects
resemble a game of chess, to be of practical importance we need to ascribe
the appropriate physical labels to the coordinates and the spaces, something
which requires skill.

\underline{{\Large Acknowledgements}}

The views expressed above have been formed over the years by many
colleagues, who include P. Halpern, the late F. Hoyle, J. Leslie, B.
Mashhoon and R. Tavakol. \ This work was supported in part by N.S.E.R.C.

\underline{{\Large References}}

\begin{itemize}
\item Barrow, J.D., 1981. \ Quant. J. Roy. Astron. Soc. \underline{22}, 388.

\item Barrow, J.D., Tipler, F.J., 1986. The Anthropic Principle. Oxford Un.
Press, New York.

\item Campbell, J.E., 1926. \ A Course of Differential Geometry. Clarendon
Press, Oxford.

\item Eddington, A.S., 1935. New Pathways in Science. Cambridge Un. Press,
Cambridge.

\item Eddington, A.S., 1939. The Philosophy of Science. Cambridge Un. Press,
Cambridge.

\item Green, M.B., Schwarz, J.H., Witten, E., 1987. Superstring Theory.
Cambridge Un. Press, Cambridge.

\item Gubser, S.S., Lykken, J.D., 2004. Strings, Branes and Extra
Dimensions. World Scientific, Singapore.

\item Halpern, P., 2004. The Great Beyond: Higher Dimensions, Parallel
Universes, and the Extraordinary Search for a Theory of Everything. J.
Wiley, Hoboken, New Jersey.

\item Kilmister, C.W., 1994. Eddinton's Search for a Fundamental Theory.
Cambridege University Press, Cambridge.

\item McCrea, W.H., Rees, M.J. (eds.), 1983. Phil. Trans. Roy. Soc. (London)
A \underline{310}, 209.

\item Petley, B.W., 1985. The Fundamental Constants and the Frontier of
Measurement. Hilger, Bristol.

\item Price, K., French, S. (eds.), 2004. Arthur Stanley Eddington:
Interdisiplinary Perspectives. Centre for Research in the Arts, Humanities
and Social Sciences (10-11 March), Cambridge.

\item Seahra, S.S., Wesson, P.S., 2003. Class Quant. Grav. \underline{20},
1321.

\item Szabo, R.J., 2004. An Introduction to String Theory and D-Brane
Dynamics. World Scientific, Singapore.

\item Wesson, P.S., 1978. Cosmology and Geophysics. Hilger/Oxford Un. Press,
New York.

\item Wesson, P.S., 1992. Space Science Reviews, \underline{59}, 365.

\item Wesson, P.S., 1999. Space, Time, Matter. World Scientific Publishing
Corporation, Singapore.

\item Wesson, P.S., 2006. Five-Dimensional Physics. World Scientific,
Singapore.

\item West, P., 1986. Introduction to Supersymmetry and Supergravity. World
Scientific, Singapore.
\end{itemize}

\end{document}